\documentclass[reprint,amsmath,amssymb,aps]{revtex4-2}

\usepackage{graphicx}
\usepackage{dcolumn}
\usepackage{bm}
\usepackage{physics}
\usepackage{mathtools}
\usepackage{relsize}
\usepackage{bigints}
\usepackage{pgfplots}
\usetikzlibrary{spy}
\usepackage{helvet}
\usepackage[colorlinks=true, allcolors=blue]{hyperref}
\usepackage{academicons}
\usepackage{blindtext}
\usepackage{subfigure}
\usepackage{amsmath}
\usepackage{amsbsy}
\usepackage{lipsum}
\usepackage{color}
\usepackage{tikz,xcolor,hyperref}
\usepackage{placeins}

\begin{document}

\preprint{APS/123-QED}

\title{Melting of colloidal crystal in a two-dimensional periodic substrate: Switch from a single 
crossover to two-stage melting}

\author{Akhilesh M P} 
\email{p20210003@goa.bits-pilani.ac.in,\\
akhileshmadathil@gmail.com}
\author{Toby Joseph} 
\email{toby@goa.bits-pilani.ac.in}
\affiliation{Department of Physics, BITS-Pilani, K K Birla Goa Campus, Zuarinagar, Goa-403726, India}

\date{\today}

\begin{abstract}

The melting transitions of a colloidal lattice confined to a two-dimensional ($2D$) periodic substrate
of square symmetry are studied using Monte Carlo simulations. When the strengths of interparticle
and particle-substrate interactions are comparable, the incommensurate nature of square and triangular 
ordering leads to the formation of a partially pinned solid with only one of the smallest {\bf G} 
vectors of the substrate present. This low-temperature phase has true long-range order.
By varying the lattice parameter of the substrate while keeping the filling fraction constant, 
it is seen that the transition from this low-temperature solid to a high-temperature 
modulated liquid phase can happen via either a single crossover transition or by a two-stage melting
process. The transitions are found to be second-order in nature when the lattice parameter is $d \lesssim 9 \lambda$,
as confirmed by the finite-size scaling behavior of the specific heat. For the two-stage melting scenario,
the intermediate phase is found to be hexatic. The transitions observed in this work
are different from the predictions of the KTHNY theory. The study reveals how constraints from substrate
periodicity can fundamentally alter melting dynamics, offering insights into the design of tunable colloidal
systems and advancing the understanding of phase transitions in two-dimensional particle systems.
\end{abstract}

\maketitle
\section {Introduction}

In $2D$, understanding the rich interplay of order, fluctuations, and topology is crucial in predicting
the phase behavior of solids under various system conditions. At finite temperatures, in the crystalline phase,
the long-wavelength fluctuations destroy the long-range positional order of the systems in $2D$. Thus, no
long-range positional order can exist in the $2D$ crystalline phase according to the Mermin-Wagner theorem
\cite{merminJMP}. A comprehensive theory of $2D$ melting based on topological defects was developed by Kosterlitz,
Thouless, Halperin, Nelson, and Young, called KTHNY theory\cite{kost, kost_2, halp, halp_2, young}. Several experimental\cite{widom,angel,beek,deut,han,horn,keim,kusn,maret,peng,rosen,scho,von,brink,zahn},
theoretical\cite{fisher,ram,chui,klein_1,klein_2,klein_3}, and simulation studies\cite{frenk,duda,krauth,lin,prest,saito,str,stran,tobo,udin,webm,wier,gribo,abra,tox,shiba,chen,marcus} 
have been carried out to understand the melting transition of solids in $2D$. Although some of the experimental
results show conformity to the predictions of the KTHNY theory \cite{widom,deut,horn,keim,kusn,maret,rosen,scho,
von,brink,zahn}, others differ in the number and nature of transitions \cite{angel,beek,han,peng}.  

According to the KTHNY melting scenario, the transitions are mediated by 
two types of topological defects: dislocations and disclinations. A dislocation can be described by
a pair of five-fold and seven-fold oriented sites, a disclination by an isolated five-fold or seven-fold
oriented site in a triangular lattice structure. KTHNY melting scheme consists of two stages (i) a transition 
from a crystalline phase to an intermediate thermodynamic phase termed, hexatic phase mediated by unbinding of
bound dislocations, and (ii) a transition from a hexatic phase to an isotropic liquid phase, mediated by unbinding
of bound disclinations. In the crystalline phase, the system has quasi-long-range positional order and long-range
bond orientational order. But in the hexatic phase, the positional order is short-range, but retains a
quasi-long-range orientational order. The existence of the hexatic phase has been confirmed in several
experimental and simulation studies\cite{han,tag,udin,som,zoll,fern,jast}. One of the key factors that 
determines whether the KTHNY scenario manifests is the core energy of the dislocation \cite{chui,saito,klein_2,klein_3}.
It is observed that when the free dislocation core energy of the system is small (less than $2.84\:k_BT$),
then the system would follow the grain boundary melting that preempts the formation of the hexatic phase\cite{yong}.

A related problem that has been of interest is the study of $2D$ particle systems in the presence of
periodic substrates. The effect of the periodic substrate can crucially depend on the symmetry of the substrate
as well as the commensurate or incommensurate nature of the substrate. In the presence of a periodic substrate 
with $1D$ modulation, the studies reveal that the system can undergo a freezing transition followed by a melting transition
as the substrate strength is increased\cite{we,le,bech_1, strep}. For periodic substrates with triangular
symmetry, Nelson et al. have predicted the existence of a commensurate solid, a floating solid, and a modulated
liquid depending upon commensuration between lattice and substrate as well as temperature\cite{nels}. In the weak
substrate limit, the melting of a Xe monolayer adsorbed on Ag identifies the presence of a hexatic phase \cite{grei},
which indicates the feature of $2D$ melting. The two-stage melting with the presence of an intermediate phase (with
short-range positional order, but solid-like) has been observed in a monolayer of Ar adsorbed on a graphite substrate
\cite{zhang}. However, in an experimental study of an incommensurate monolayer of solid adsorbed on a $2D$ substrate,
a single peak in the specific heat has been observed that corroborates a first-order phase transition in the system
\cite{dash}.

Studies have been carried out on $2D$ particle systems in periodic square symmetric substrates 
(the ratio of the number of particles to the number of substrate minima) \cite{nels,neuh,toby_1}.
Nelson et al. predicted that an Ising-like transition replaces the disclination unbinding transition in the usual
KTHNY scenario of phase transition \cite{nels}. Using Monte-Carlo simulations, Toby et al.\cite{toby_1} have studied
the phases and phase transition of the vortex lattice in a $2D$ periodic substrate of square symmetry when the strengths
of the inter-vortex interaction and the vortex-substrate interactions are comparable. For a filling fraction (number of
vortices per unit cell of square substrate) equal to one, a low-temperature partially pinned phase was observed. As the
temperature is increased, this phase melts to a modulated square liquid via a second-order transition. A similar phase 
has been identified by T. Neuhaus et al. (referred to as the rhombic phase in that work), in a system of hard disks in a
square symmetric substrate \cite{neuh}, by employing the Monte-Carlo simulation and density functional theory(DFT). For 
high substrate strength, a single second-order transition is seen from the rhombic phase to the modulated phase as the 
packing fraction is decreased. At low substrate strengths, they find three separate phases with the rhombic phase giving
way to triangular order at high packing fractions.

In the current work, we propose to study a system of colloidal particles interacting via a screened Coulomb potential
in the presence of a square-symmetric potential for filling fraction equal to one. The key parameters concerning which we study
the phases of the system are the size of the square substrate (relative to the screening length) and the strength of
the substrate. Though the partially pinned solid and its continuous transition to a modulated liquid are found, like 
in the case of the vortex lattice study \cite{toby_1}, a two-stage melting is observed as the size of the square substrate
is decreased. Both these transitions are found to be second-order in nature, as confirmed by the finite-size scaling study
of the specific heat. 

This paper is organized as follows: In section \ref{sim}, we introduce the system, the simulation method used, and the 
details of the various quantities studied in the simulations. In section \ref{disc}, the results obtained
from the study are presented, followed by a discussion of the results. In section \ref{sum}, we conclude the work, discuss
possible experimental avenues to explore, and discuss future directions.
\begin{figure}
    \centering
    \includegraphics[scale=0.6]{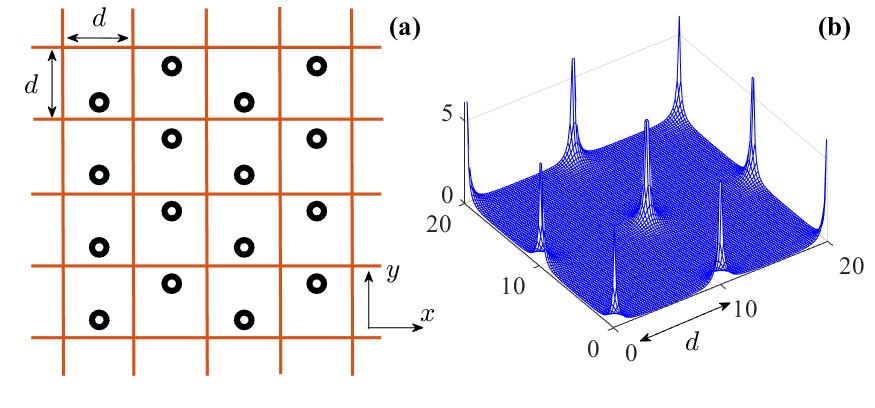}
    \caption{(a) A schematic of the partially pinned structure of the colloidal lattice with the underlying
    substrate. The black circles represent the colloidal particles, and the red lines indicate the underlying
    square symmetric substrate with spacing $d$. The pinned colloidal particles are located at the corners of
    the squares (b) The potential landscape of the substrate formed by the pinned particles which interact with
    the interstitial ones via screened Coulomb interaction.}
    \label{pps}
\end{figure}

\section {Model and Simulation}\label{sim}

The system that has been investigated encompasses a collection of $N_p$ interstitial colloidal particles 
(referred to as particles henceforth) confined to move in a $2D$ periodic square symmetric substrate. 
The substrate is effected by the interaction of these interstitial colloidal particles with a fixed set
of colloidal particles (referred to as pinned particles henceforth), located at the corners of a square
lattice of sides $d$ (see Fig.\ref{pps}). The interstitial particles interact with each other by 
screened Coulomb interaction given by
\begin{equation}
    {U}_{i}^{cc} = U_0
    \sum_{j=1\atop j\neq i}^{N_p} \frac{e^{-\frac{r_{ij}}{\lambda}}}{r_{ij}}
\end{equation}
where $U_i^{cc}$ is the energy of the $i^{th}$ particle due
to the interaction between other particles in the lattice, $r_{ij}=|{\bf r}_{i}-{\bf r}_{j}|$ is the interparticle
separation, $U_0$ is the strength of the interaction potential, and $\lambda$ is the screening length of the potential.
The interstitial particles also interact with the pinned particles via the same type of interaction, but with different
interaction strength. The energy of the $i^{th}$ particle in the lattice as a result of the interaction
with the pinned particles of the substrate is given by
\begin{equation}
    {U}_{i}^{cp} = A_sU_0
    \sum_{k=1\atop k\neq i}^{N_p} \frac{e^{-\frac{r_{ik}}{\lambda}}}{r_{ik}}
\end{equation}
where $A_s$ is a parameter that accounts for the effect of the strength of the substrate. 
The system is in contact with a thermal reservoir at temperature $T$. The Monte Carlo simulation technique that utilizes
the Metropolis-Hastings algorithm is used to study the equilibrium phase behavior of the system. It is assumed that the
pinned particles wouldn't be depinned from their respective pinned sites due to thermal fluctuations in the system.
So the dynamics of these particles can be ignored in the simulations.  

A thermodynamic quantity of interest in understanding the nature of the phase transition is the specific heat 
at constant volume $C_v$. It can be determined by the fluctuation in energy of the system by the relation
\begin{equation}
    C_v = \frac{\left< (E-\left< E \right>)^2 \right >}{N_pk_BT^2}
\end{equation}
where $E$ is the total energy of the system and $k_B$ is the Boltzmann constant. The low-temperature phase
obtained in the study is a crystal with long-range positional order. To find out the translational(positional) 
symmetry breaking point, the global translational order parameter for the two smallest reciprocal lattice vectors ${\bf G}_1$ 
and ${\bf G}_2$ of the lattice is calculated using, 
\begin{equation}
    \Psi_T = \left|\frac{1}{N_p}\sum_{i=1}^{N_p}\psi_{T_i} \right| 
\end{equation}
where $\psi_{T_i} = e^{i({\bf G}.{\bf r}_i)}$ is the local translational order parameter for the $i^{th}$ particle 
at the position ${\bf r}_i=(x,y)$. The translational order parameter can give information about the structural 
change due to the change in position of the particle during melting. The static structure factor is computed using
the vector ${\bf q}$ in the reciprocal lattice space by
\begin{equation}
    S({\bf q})=\frac{1}{N_p^2} \sum_{i,j}e^{i{\bf q}.({\bf r}_{i}-{\bf r}_{j})}
\end{equation}
To locate the point at which the orientational symmetry of the lattice breaks, the global orientational
order parameter defined by,
\begin{equation}
    \Psi_6 = \left|\frac{1}{N_p}\sum_{i=1}^{N_p}\psi_{6_i} \right| 
\end{equation}
is computed. Here $\psi_{6_i} = \sum_{j=1}^{n_i} e^{i(6\theta_{ij})}/n_i$ is the local orientational order parameter for the
$i^{th}$ particle, $\theta_{ij}$ is the angle between a bond formed by $i^{th}$ and $j^{th}$ particles and a 
reference axis, and $n_i$ is the number of nearest neighbors of the $i^{th}$ particle. The orientational order
parameter gives information about the change in bond orientational order in the lattice. 

To find the transition points accurately, the susceptibility is a useful quantity to study. It is
defined as the fluctuation of the order parameter. This method would work well in identifying the transition points 
as it has less finite-size effects inherent in correlation functions\cite{han}. The susceptibilities are given by
\begin{equation}
    \chi_a = A(\left<\psi_a^2 \right> - \left< \psi_a \right>^2)
\end{equation}
where $A$ is the area of the system and the subscript $a$ stands for $T$, $6$. $\chi_T$ measures the translational
susceptibility and $\chi_6$ measures the orientational susceptibility. The peak in susceptibility values
can be used to accurately locate the transition points, both in simulation\cite{webm,ronen} and in experiments\cite{han,ziren}.
To characterize the different
phases of the system, the orientational correlation function given by
\begin{equation}
    g_6({\bf r}) = \left< \psi^{*}_6({\bf r}_i)\psi_6({\bf r}_j)\right>
\end{equation}
has been calculated. Here, $\psi_6({\bf r})$ represents the local bond orientational order parameter of the particle at
the position ${\bf r}$.

In simulation, the following units are used: $\lambda=1$ and  $k_B T$ in units of 
$U_0\:\frac{e^{-d/\lambda}}{d}$ with $U_0=1$. The average values of the various quantities in the study
have been carried out over $5\times N_p\times10^8$ Monte Carlo steps after allowing the system to equilibrate for $N_p\times10^5$ 
Monte Carlo steps. The simulated annealing technique has been used to obtain the low-temperature structure of the system.
Additionally, it is ensured that the number of Monte Carlo steps is sufficient for the proper equilibration of the system.
Though most of our simulations are carried out for $L \times L$ system, we have ensured that the imposition of square boundaries
does not have any influence on our results. This was done by verifying our results using a rectangular-shaped boundary that is
commensurate with the triangular lattice for the same density. To understand the role of defect generation in the melting 
of the lattice, Voronoi-tessellation is performed. Using
this technique, the distribution of topological defects: isolated dislocations, disclinations, and bound dislocations,
in different phases is studied. Simulation studies are carried out for various system sizes to do finite-size scaling analysis.


\section {Results and discussion}\label{disc}
\begin{figure}
    \centering
    \includegraphics[scale=0.6]{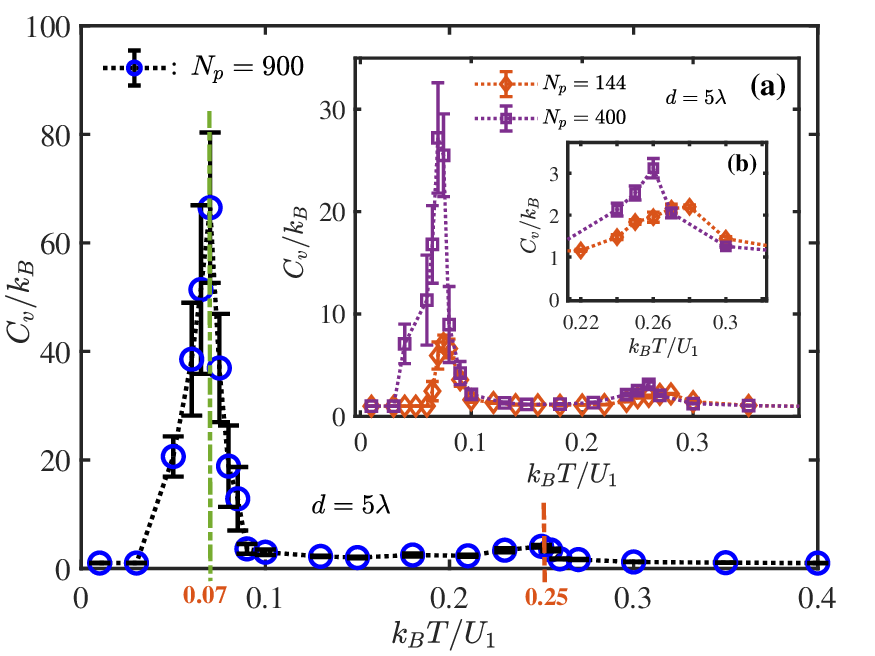}
    \caption{Main plot: The variation of specific heat $C_v$ (in units of $k_B$) as a function of $k_BT$ (in units of
    $U_1=U_0\frac{e^{-5}}{5\lambda}$) at $d=5\lambda$ for $N_p=900$. The $C_v$ has two peaks,
    one at $k_BT_{c1}=0.07\:U_1$ and the other at $k_BT_{c2}=0.25\:U_1$. These two peaks correspond to the
    phase transitions from solid to hexatic and hexatic to modulated liquid phases. The inset plot (a) shows the 
    variation of $C_v$ for the other two smaller system sizes with $N_p=144$ and $400$. For $N_p=144$, the 
    solid to hexatic transition happens at $k_BT_{c1}=0.075\:U_1$ and the hexatic to liquid transition at $k_BT_{c2}=0.28\:U_1$
    Similarly, for $N_p=400$, the solid to hexatic transition occurs at $k_BT_{c1}=0.07\:U_1$ and the hexatic to liquid
    transition at $k_BT_{c2}=0.26\:U_1$. Inset plot (b) shows the enlarged portion of the hexatic-liquid transition
    for $N_p=144$ and $N_p=400$. There is a clear shift in both the transition temperatures with system size as well
    as an increase in the $C_v$ values at both transitions. Here, the error bars indicate the standard deviation of the mean
    , and the dotted lines shown are guides to the eye.}
    \label{Cv_900}
\end{figure}
\begin{figure}
    \centering
    \includegraphics[scale=0.6]{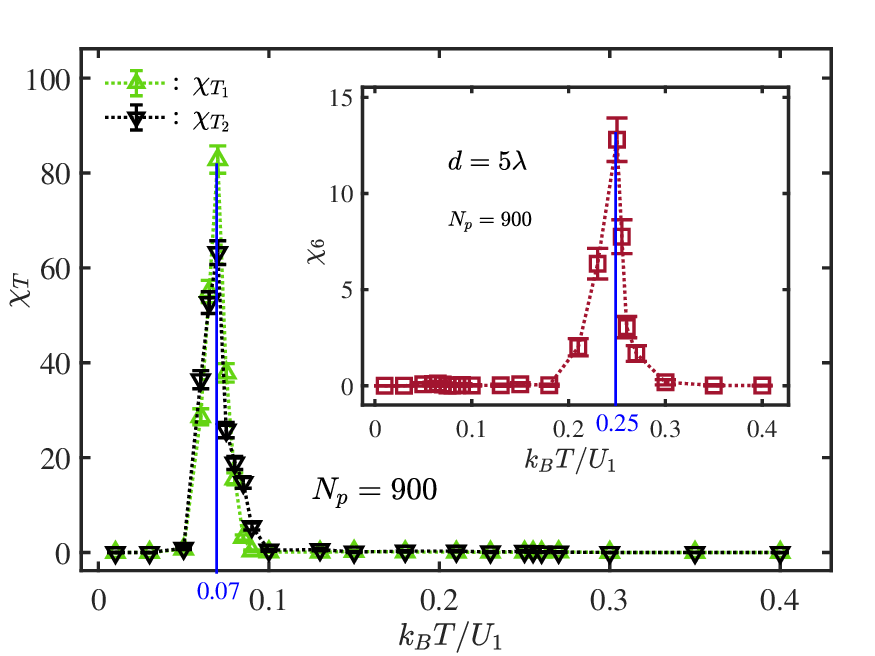}
    \caption{Main plot: Variation of the translational susceptibilities $\chi_{T1}$ and $\chi_{T2}$ as a function of $k_BT$ (in
    units of $U_1=U_0\:\frac{e^{-5}}{5\lambda}$) for $N_p=900$ at $d=5\lambda$. The inset plot shows the
    variation of orientational susceptibility $\chi_6$ as a function of $k_BT$. Both the plots accurately locate 
    the transition temperatures $k_BT_{c1}=0.07\:U_1$ and $k_BT_{c2}=0.25\:U_1$
    marked by a solid blue line, using the peaks of $\chi_T$'s and $\chi_6$. Error bars represent the standard deviation
    of the mean, and the dotted lines indicate guides to the eye. }
    \label{chi_900}
\end{figure}
\begin{figure}
    \centering
    \includegraphics[scale=0.54]{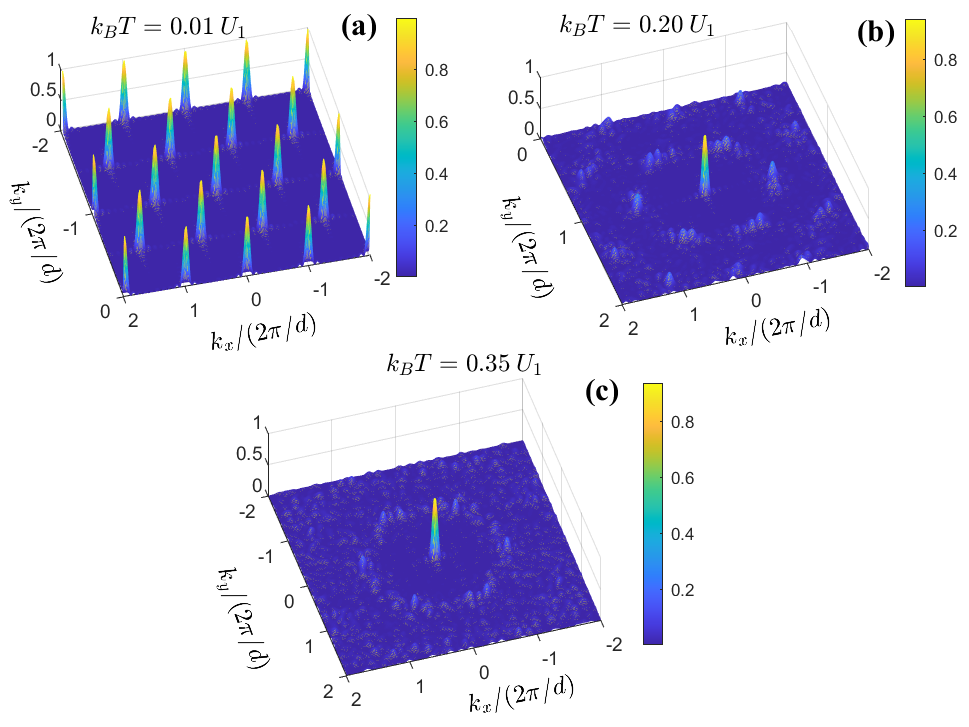}
    \caption{Structure factor in different phases for the case $d=5\lambda$.
    (a) The partially pinned solid phase with true long-range positional and orientational orders
    at $k_BT=0.01\:U_1$. (b) The intermediate hexatic phase with short-range positional order and 
    quasi-long-range orientational order at $k_BT=0.20\:U_1$. (c) The liquid phase with short-range 
    positional and orientational orders at $k_BT=0.35\:U_1$.}
    \label{surf}
\end{figure}

The simulation studies have been carried out for filling fraction, $n = 1$, by changing both the lattice spacing
$d$ and substrate strength, $A_s$. For $d = 5\lambda$ and $A_s = 10^{-4}$, the low temperature phase is the partially pinned 
solid (rhombic structure) with an equal distance of separation $\approx 0.13\:d$ from the minima on either side of the square cell,
similar to the one observed in vortex lattice simulation \cite{toby_1} and for the hard sphere system \cite{neuh}.
Remarkably, a two-stage melting is observed in this case.
The variation of $C_v$ with temperature for this case is shown in Fig.\ref{Cv_900}. Two distinct peaks, one at
$k_BT_{c1}=0.07\:U_1$ and the other at $k_BT_{c2}=0.25\:U_1$ are seen, indicating the possible occurrence of 
two phase transitions. Here, $U_1=U_0\frac{e^{-5}}{5\lambda}$ is the energy
scale corresponding to the inter-particle interaction for the present case. Different system sizes ($L\times L = 144,
\:196,\: 256,\: 400,\: 676,$ and $900$, $L$ being the linear size of the system) were studied to ensure the
robustness of the results and to check the presence of scaling. It has been observed that the peaks in $C_v$ shift
towards lower temperatures as the size of the system is increased. Two peaks in the $C_v$ plot indicate the presence
of an intermediate phase between the solid and the modulated liquid. 
\begin{figure}
    \centering
    \includegraphics[scale=0.6]{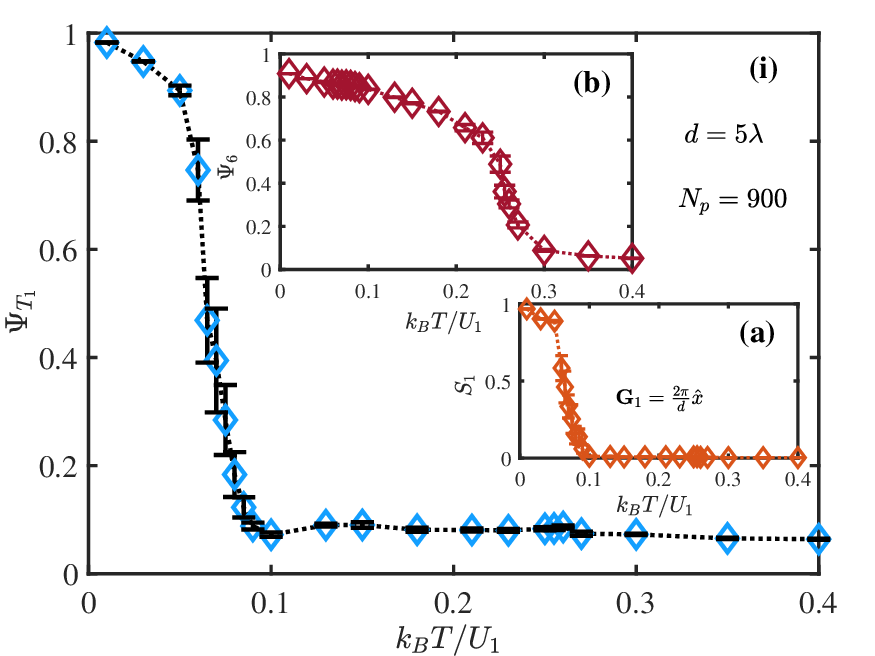}
    \includegraphics[scale=0.6]{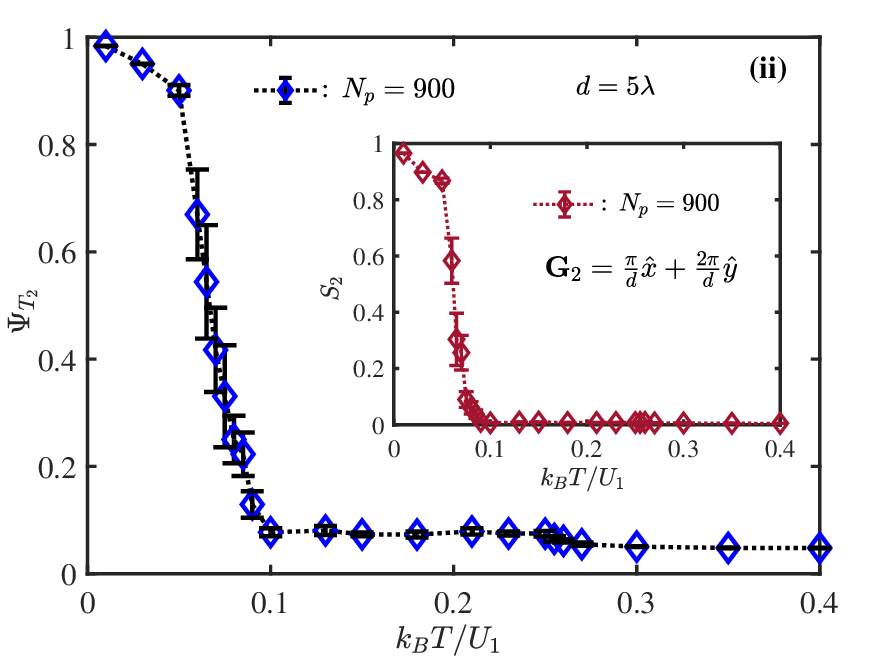}
    \caption{(i) Variation of translational order parameter $\Psi_{T1}$ corresponding to ${\bf G_1}$ as a function
    of $k_BT$ (in units of $U_1=U_0\:\frac{e^{-5}}{5\lambda}$) for $N_p=900$ at $d=5\lambda$. The corresponding structure factor
    $S_1$ for ${\bf G_1}$ is shown in the inset plot (a). Here, both $\Psi_{T1}$ and $S_1$ show a continuous drop around the
    transition temperature $k_BT_{c1}$. Inset plot (b) shows the variation of the orientational order parameter $\Psi_6$ with $k_BT$. It
    has a continuous drop around the second transition temperature $k_BT_{c2}$.  (ii) Variation of translational
    order parameter $\Psi_{T2}$ as a function of $k_BT$ for $N_p=900$ at $d=5\lambda$. The corresponding variation of $S_2$
    for the reciprocal lattice vector ${\bf G_2}$ is shown in the inset plot. $\Psi_{T2}$ and $S_2$ also drop continuously
    around the transition temperature $k_BT_{c1}$. The dotted line indicates the guide to the eye, and the error bars are the 
    standard deviation of the mean.}
    \label{psi6_900}
\end{figure}
The order parameters $\Psi_{T_1}, \Psi_{T_2}$ and $\Psi_{6}$ as well as the corresponding susceptibilities were computed
(see Fig.\ref{psi6_900}, \ref{chi_900}) to gain insight into the nature of the phases involved. 
The translational and bond orientational orders drop smoothly around two different temperatures.
Since the variations of $\Psi_{T_1}, \Psi_{T_2}$ and $\Psi_{6}$ with $k_BT$ are inconclusive to locate the transition points,
the susceptibilities of the order parameters have been computed. 
The corresponding susceptibilities show sharp peaks at two distinct temperatures, which are aligned with the location of the observed 
peaks in $C_v$ at all system sizes studied. These findings indicate that the intermediate phase is a hexatic phase, 
with no long-range translational order but with bond orientational order being present. The static structure factor
for various phases is shown in Fig.\ref{surf} and clearly indicates the solid, intermediate, and liquid phases at
various temperatures. The change in the peak value of the structure factor corresponding to the reciprocal 
lattice vectors ${\bf G_1}$ and ${\bf G_2}$ with temperature is plotted in Fig.\ref{psi6_900} (see insets).
These two display a smooth drop near the temperature corresponding to the loss of positional order, as in the case of
the translational order parameter.


\begin{figure}
    \centering
    \includegraphics[scale=0.6]{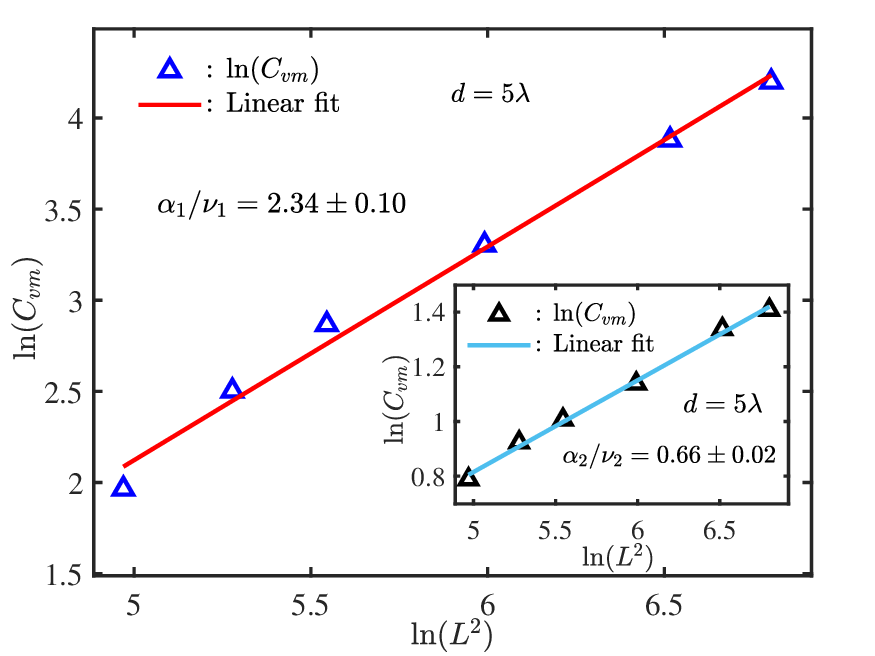}
    \caption{Main plot: Scaling behavior of the first transition peak $C_{vm}$ of the specific heat as a 
    function of system size $L^2$ in the log-log scale for $d=5\lambda$. The inset plot shows the scaling
    behavior of the second transition peak $C_{vm}$ of the specific heat with system size $L^2$ in the log-log
    scale for $d=5\lambda$. The scaling exponents are $\alpha_1/\nu_1=2.34\pm0.10$ and $\alpha_2/\nu_2=0.66\pm0.02$ 
    respectively for the first and second transitions. The triangles in the plots represent simulation points, 
    and the solid line indicates the corresponding linear fit.}
    \label{scale}
\end{figure}


A finite-size scaling analysis of the peaks in specific heat $C_v$ for $ d=5\lambda$ (see Fig.\ref{scale}) has been carried
out in the system to confirm the nature of the phase transitions. As seen Fig.\ref{Cv_900}, peaks in $C_v$ become sharper
when the size of the system is increased and the transition temperature reduces. The peak value of specific heat, $C_{vm}$,
is found to scale as $L^{\alpha/\nu}$, where $L=\sqrt{N_p}$, at both the transitions. The scaling exponents for the 
solid-hexatic and the hexatic-liquid transitions are found to have values $\alpha_1/\nu_1 = 2.34\pm0.10$ and 
$\alpha_2/\nu_2 = 0.66\pm0.02$, respectively. This analysis points again to the fact that the transitions observed
in this case are second-order. In the KTHNY type of melting, one expects a peak in specific heat, but it is not expected
to scale with system size. For example, an experimental study carried out on superparamagnetic polystyrene beads
confined in $2D$ shows a single peak in specific heat $C_v$ as a function of temperature \cite{sven}.
The observed peak in $C_v$ in the hexatic phase in that work indicates the maximum increase in the value of 
isolated dislocations.
\begin{figure}
    \centering
    \includegraphics[scale=0.62]{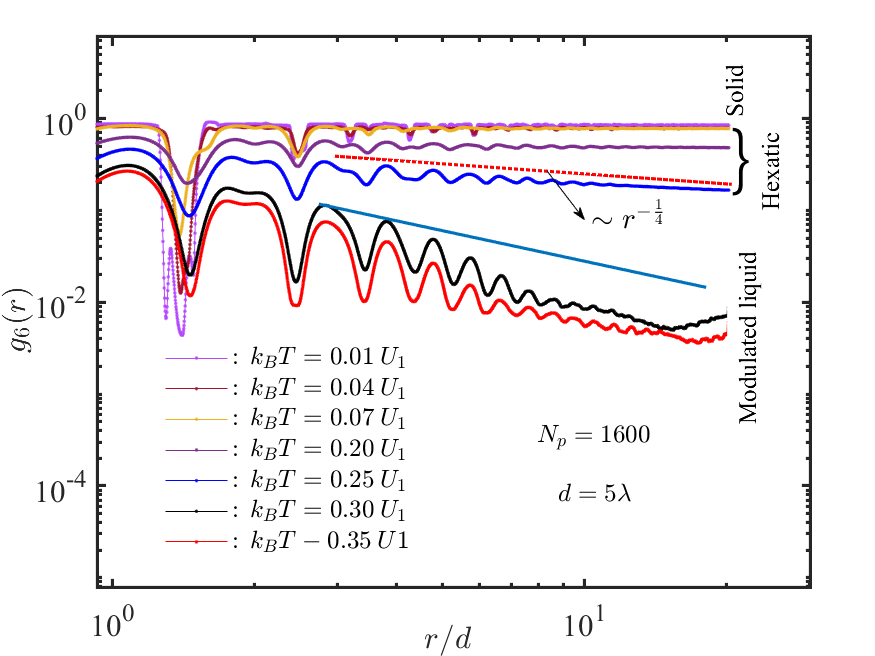}
    \caption{The variation of orientational correlation function $g_6(r)$ as a function of radial distance
    $r$ in log-log scale for different values of $k_BT$ for $d=5\lambda$, $Np=1600$. Three different phases
    are marked in the plot based on the nature of $g_6(r)$. In the hexatic phase, close to the hexatic-modulated 
    liquid transition temperature, the correlation exponent $\eta_6(T)$ has a value close to $1/4$ (slope of the
    red dotted line is $0.243 \pm 0.012$). This is aligned with the prediction of the KTHNY theory. In the modulated liquid phase, 
    $g_6$ has faster algebraic decay, indicating the absence of any long-range order.}
    \label{ori_cor}
\end{figure}

Since both transitions seen for $d=5\lambda$ are found to be second-order, which is novel and contrary to what
the KTHNY theory predicts, it would be interesting to look at the correlation functions and defect structure in various
phases for comparison. Orientational correlation function $g_6(r)$ (calculated for $N_p=1600$) at different temperatures 
are shown Fig. \ref{ori_cor}. $g_6(r)$  exhibits three distinct behaviors with $r$, depending on which of 
the three phases the system is in: solid, hexatic, or liquid. There is long-range order in the solid phase,
algebraic decay in the hexatic phase, and exponential decay in the liquid phase. The nature of the exponent for
the algebraic decay, $\eta_6(T)$, is such that it starts from low values at the transition point from solid to
hexatic and reaches a maximum value close to $\frac{1}{4}$ just before melting to the modulated liquid phase.
One has an interesting situation wherein the system melts via two-stage melting as predicted by KTHNY theory, 
but the nature of the observed transitions is different from the KTHNY prediction. However, the orientational exponent
$\eta_6(T)$ at the hexatic-liquid transition temperature is found to be close to $\frac{1}{4}$ ($0.243\pm 0.012$). This value of $\eta_6(T)$
consistent with the value predicted by the KTHNY theory.
\begin{figure}
    \centering
    \includegraphics[scale=0.645]{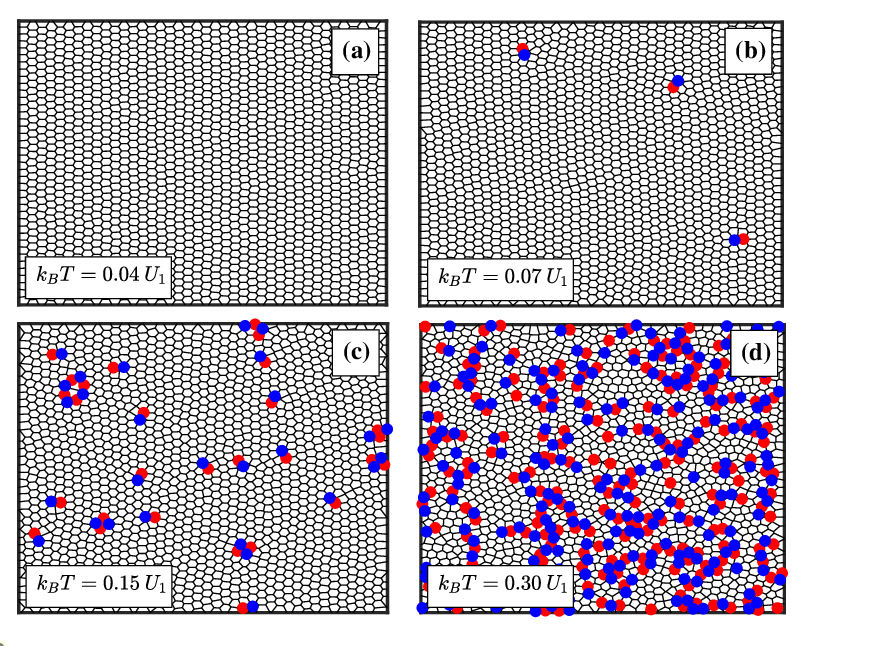}
    \caption{Defect generation during the two-stage melting of the colloidal lattice ($d=5\lambda$ and system
    size is $N_p = 900$). The red and blue dots indicate five-fold and seven-fold coordinated
    sites, respectively. The uncolored sites represent six-fold coordinated sites. 
    (a) The crystalline phase with no defects at $k_BT=0.04\:U_1$. (b) Unbound dislocations (or bound disclinations) 
    are found close to the solid-hexatic transition point ($k_BT=0.07\:U_1$). (c) In the hexatic phase at $k_BT=0.15\:U_1$, dislocations
    concentration increases considerably, and a fraction of them are found in bound form. A chain of defects formed by alternating
    coordination numbers can also be seen. (d) Defects proliferate in the liquid phase at $k_BT=0.30\:U_1$, with many free 
    disinclinations appearing along with chains of defects.}
    \label{def} 
\end{figure}

The results of the defect studies are shown in Fig.\ref{def}. In the solid phase at low temperatures, the
system is free of defects (Fig.\ref{def} (a)). This is expected as the system has true long-range order in this phase. 
Close to the solid to hexatic transition, free dislocations appear (Fig.\ref{def} (b)). In the hexatic regime,
dislocations increase with temperature, some of which tend to form bound pairs, but many remain unbound (Fig.\ref{def} (c)).
The unbound dislocations (or bound disclinations) are the reason for the loss of long-range bond orientational order
in this phase. In the modulated liquid phase, the defect concentration is high, and seems to have a combination of
free disclinations, a chain of defects, free dislocations, and a few bound dislocations (Fig.\ref{def} (d)). The defect
structure study shows that though there are similarities between the behavior observed and the KTHNY predictions for the
homogeneous case, there are crucial differences. The intermediate phase observed is similar to the hexatic phase, borne out by
both the defect structure and the $g_6(r)$ behavior. The transition to the modulated liquid phase seems to be
driven by bound disclinations unbinding. The low-temperature crystal phase is free from defects, and its melting to the hexatic
phase is via a conventional, symmetry-breaking second-order transition.

\begin{figure}
    \centering
    \includegraphics[scale=0.63]{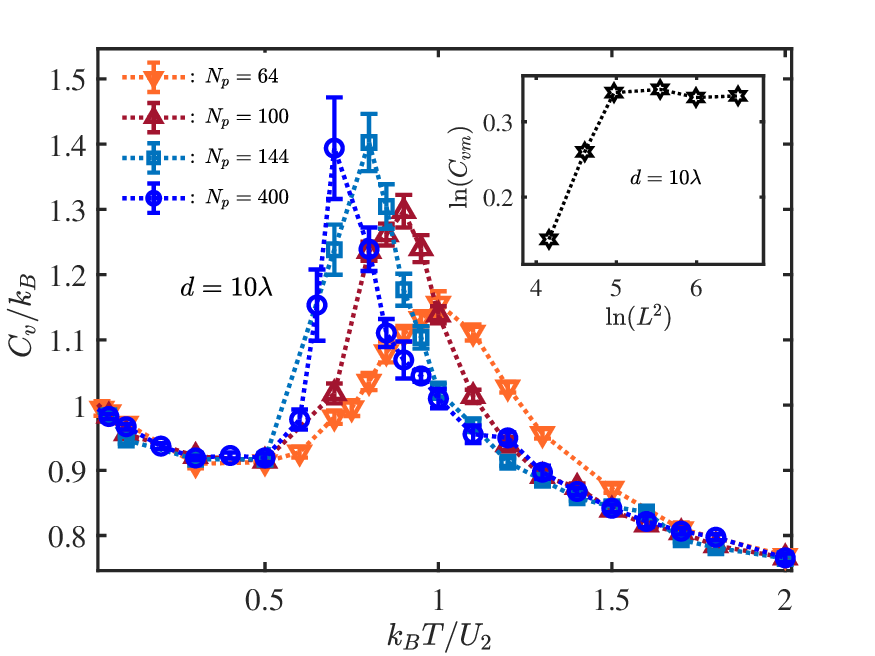}
    \caption{The variation of specific heat $C_v$ (in units of $k_B$) as a function of $k_BT$ (in units of $U_2=U_0\:\frac{e^{-10}}{10\lambda}$) for $N_p=64,100, 144$, and $400$ at $d=10\lambda$. A single peak in the specific heat is observed, and the location of 
    the peak shifts to low temperatures with increasing system size. The inset plot shows the variation of the peak value of 
    specific heat ($C_{vm}$) as a function of system size $L^2$ in the log-log scale for $d=10\lambda$. The dotted lines are guides to the eye, and the error bars are the standard deviation of the mean.}
    \label{cv_d10}
\end{figure}
To explore the effect of the substrate lattice parameter, $d$, on phases and phase transitions, simulations
were carried out for $d = 10\lambda$ with the filling fraction kept at $n = 1$.
For $d = 10\lambda$ and $A_s = 10^{-4}$, the ground state is found to be the partially pinned solid in this case
as well, with an equal separation distance of $\approx 0.23\:d$ from the minima on either side of the square cell. 
This phase has true long-range translational and orientational orders (See Fig.\ref{pps}). This phase undergoes a
transition to a modulated liquid phase as the temperature is increased. The variation of specific heat $C_v$ as a function
of $k_BT$ for $Np=400, 144, 100$ and $100$ at $d=10\lambda$ is shown in Fig.\ref{cv_d10}. A single peak in $C_v$ is observed.
The location of the peak saturates to a value close to $k_BT_c= 0.7\:U_2$ as the system size is increased.
The locations of the peak in $C_v$ for the system sizes $N_p=400, 144, 100$ and $64$ are respectively at
$k_BT_c= 0.7\:U_2,\:0.8\:U_2,\:0.9\:U_2$ and $0.95\:U_2$, where $U_2 = U_0\:\frac{e^{-10}}{10\lambda}$, is the energy scale
of inter-particle interaction. The peak value of $C_v$ is seen to saturate with the size of the system (see the inset of
Fig.\ref{cv_d10}). 
\begin{figure}
    \centering
    \includegraphics[scale=0.6]{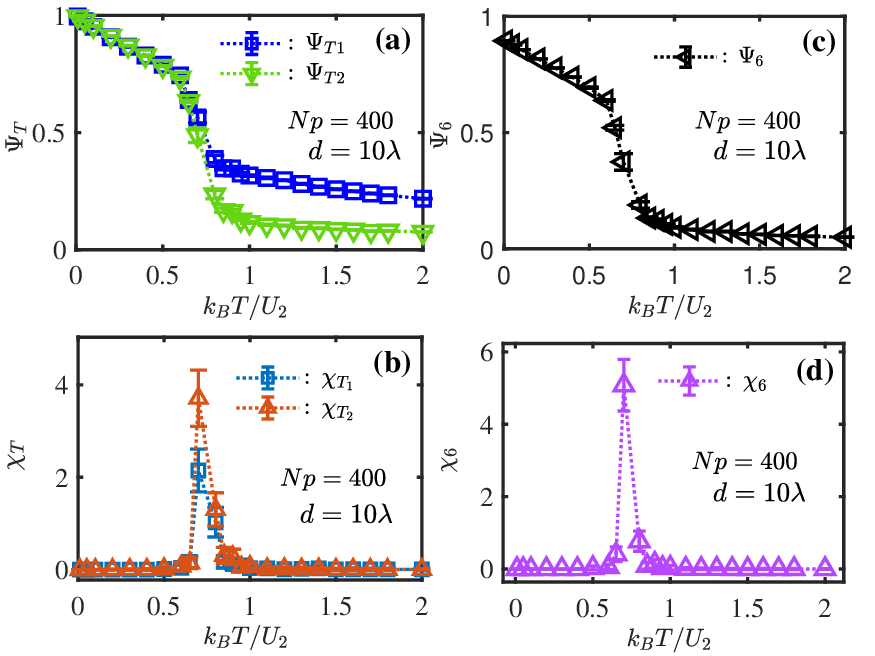}
    \caption{(a) Variation of the translational order parameters $\Psi_{T1}$ (corresponding to ${\bf G_1}$) and
    $\Psi_{T2}$ (corresponding to ${\bf G_2}$) as a function of $k_BT$ (in units of $U_2=U_0\:\frac{e^{-10}}{10\lambda}$)
    for $N_p=400$ at $d=10\lambda$. (b) The corresponding susceptibilities ($\chi_{T_1}$ \& $\chi_{T_2}$). 
    Plots (c) \& (d) show the variation of the orientational order parameter $\Psi_6$ and its corresponding 
    $\chi_6$ as a function of $k_BT$. The behavior seen clearly indicates a single transition at which the
    solid melts to a modulated liquid. Here, symbols are the simulation points, and the dotted lines are 
    guides to the eye. Error bars shown in the plots are the standard deviation of the mean.}
    \label{psi_d10}
\end{figure}

The translational ($\Psi_T$) and the bond orientational ($\Psi_6$) order parameters
show a continuous drop near the transition temperature. The order parameters for $N_p=400$ at $d=10\lambda$
are shown in Fig.\ref{psi_d10}. The translational order parameter ($\Psi_T$) is computed for the two smallest
reciprocal lattice vectors ${\bf G_1} = \frac{2\pi}{d}\hat{x}$ and ${\bf G_2} = \frac{\pi}{d}\hat{x}\: +\:\frac{2\pi}{d}\hat{y}$
of the partially pinned solid. Since the order parameter calculations are inconclusive in locating the transition points,
the susceptibility calculation has been carried out from the fluctuations of the order parameters to locate the points where the 
phase transition happens. The peak in susceptibilities, $\chi_6$, $\chi_{T_1}$, and $\chi_{T_2}$, is found to be located 
at the same temperature to within the resolution that simulations could provide. This temperature coincides with the peak in
$C_v$ for the large system size limit. These results point to a scenario where the low-temperature solid phase transitions to
a modulated liquid via a continuous crossover. This result is different from the observations in the previous work on the vortex lattice melting \cite{toby_1}, where a continuous transition was observed.

To address the question of how the nature of the transition changes from two-stage melting to a single crossover,
the transitions have been studied for a range of values of $d$ with the value of $A_s$ fixed at $10^{-4}$. We find
that for the value of $d \lesssim 9 \lambda$, there are two second-order transitions similar to $d = 5\lambda$ case
presented above. These two transitions come close and merge to a single crossover for $d\approx 9\lambda$. This seems
to indicate that for smaller values of $d$, the system behaves similarly to the homogeneous system, which also exhibits
two-stage melting. In fact, it can be argued that as $d$ is decreased, the effect of the substrate becomes weaker as
follows. The interaction of a colloidal particle with a pinned particle located at the corner of the square cell is given by
\begin{equation}
    U^{cp} = A_sU_0\:\frac{e^{-\frac{d}{\sqrt{2}\lambda}}}{\frac{d}{\sqrt{2}}}
    \label{eq1}
\end{equation}
Similarly, the interaction between two colloidal particles separated by a lattice spacing, $d$, is given by
\begin{equation}
    U^{cc}=U_0\:\frac{e^{-\frac{d}{\lambda}}}{d}
    \label{eq2}
\end{equation}
The ratio $\frac{U^{cp}}{U^{cc}}$ gives an estimate of the strength of the substrate relative to the inter-particle
interaction and is given by
\begin{equation}
    \frac{U^{cp}}{U^{cc}}=\sqrt{2}\:A_s e^{d(1 - \frac{1}{\sqrt{2}})} \;.
    \label{eq3}
\end{equation}
This ratio decreases with $d$, indicating the effect of the substrate is reduced at smaller $d$ values. Consistent
with this observation, we find that for large enough $d$, the ground state has square symmetry, implying that
the substrate dictates the symmetry of the ground state.

To cross-check the above hypothesis, simulations were done with $d$ kept fixed at $5 \lambda$ and $A_s$
varied. It is found that for larger values of $A_s$ ($A_s \gtrsim 0.001$), a single transition is
observed where both $\psi_T$ and $\psi_6$ go to zero simultaneously. The specific heat shows a peak but shows
no scaling behavior with system size, implying a crossover. For large enough values of $A_s$ ($A_s \gtrsim 0.1$),
the ground state has square symmetry and no transition is observed when the temperature is increased. These results
are consistent with the results discussed above with varying $d$ values. In the limit when the substrate strength, 
$A_s$ goes to zero, our results match the two-stage melting of the homogeneous case, observed by 
Yong Chen et al. \cite{yong} in their studies. The results based on the order parameter, susceptibility, 
and orientational correlation function calculations show agreement with the KTHNY theory predictions, 
but the observed defect structure was complex compared to the predictions of the KTHNY theory. 

\section{Summary}\label{sum}
The equilibrium phase behavior of a colloidal lattice in a $2D$ periodic substrate of square symmetry has been studied 
using the Monte-Carlo simulation for filling fraction $n=1$. Earlier studies on similar systems have shown 
that when the particle-substrate interaction becomes comparable to the inter-particle interaction,
the system forms a partially pinned structure with only one of the basic periodicities associated with the substrate present.
\cite{tob_phys, toby_1,neuh}. This phase melts into a modulated liquid phase via a single second-order transition.
The focus of the present work is to study how the colloidal system behaves as the substrate parameter, $d$, is varied 
with the screening length $\lambda$ fixed. The key findings of this study are the following:\\

{\bf (i)} For values of $d \gtrapprox 9 \lambda$, there is a single transition or crossover from a
solid to a modulated liquid. This is unlike what is observed in the hard-sphere and vortex
systems.\\

{\bf (ii)} For lower values of $d$, there is a two-stage melting from the low-temperature crystal to
a modulated liquid, with an intermediate hexatic phase being present. Both these transitions are found
to be second-order in nature. These transitions are not of Ising-type as is predicted by Nelson et al. \cite{nels}.\\

{\bf (iii)} On increasing the substrate strength for a fixed value of $d$, the two-stage melting is replaced by
a single crossover from solid to modulated liquid, much like the observation in point (i) above. Both these
behaviors can be attributed to the increasing influence of the lattice as $d$ or $A_s$ is increased.\\

{\bf (iv)} If the substrate strength is too strong (for example, $A_s > 0.1$ for $d = 5 \lambda$), the
ground state has square symmetry, and no transition is observed with increasing temperature. \\

{\bf (v)} The results from the defect study confirm the presence of an intermediate hexatic phase with 
free dislocations.  \\

The results obtained in this work give insights into the problem of the melting transitions of a $2D$ solid when
the strength of the underlying substrate is weak for filling fraction $n=1$. Even though the system undergoes 
two continuous transitions for low substrate strengths, these transitions are different from the continuous
transitions predicted by KTHNY theory of $2D$ melting. During the transition from the solid to the hexatic phase, the
system loses long-range positional order. In the second transition, the quasi-long-range orientational order is lost.
Though this transition is not a KT-type transition, the exponent $\eta_6(T)$ corresponding to the algebraic decay of
orientational correlation function is close to the value of ${\frac{1}{4}}$ as predicted by the KTHNY theory.
At both transitions $C_v$ shows clear scaling with system size, confirming that the transitions are of second order.
Another related $2D$ system where multiple transitions have been observed is the generalized $XY$ model
with higher order nematic-like couplings \cite{zuko}. 

The focus of the current study has been to understand the different phases and phase transitions of the colloidal system
in a square periodic substrate for the case of filling, $n = 1$. Both the substrate strength, $A_s$, and lattice
parameter, $d$, have been varied to see their influence on the phases. A natural extension of the work would be to study the
system for other fillings. The observation of two continuous second-order transitions in a 2D particle system is novel.
Its possible connection to the generalized $XY$ model, where multiple continuous transitions have been seen, is intriguing. 
It is to be noted here that even though the current work is carried out for colloidal systems, the physics behind
the phenomenon observed in this work would be relevant for other systems like superconducting vortices, particles adsorbed
on a metallic surface, and other relevant systems.

\acknowledgments
The authors would like to acknowledge the financial support under the DST-SERB 
Grant No: CRG/2020/003646.

\bibliographystyle{apsrev4-1} 

%

\end{document}